\documentclass[aps,prl,reprint,superscriptaddress,floatfix]{revtex4-2}

\usepackage[tight]{units}
\usepackage{textcomp}
\usepackage{gensymb} 
\usepackage{amsmath}
\usepackage{amssymb}
\usepackage{amsfonts}
\usepackage{graphicx}
\usepackage{bm} 
\usepackage{xcolor} 

\begin{document}

\title{Anisotropic Paramagnetic Peak Effect in Reversible Magnetization of Crystalline Miassite Superconductor $\text{Rh}_{17}\text{S}_{15}$}

\author{Ruslan Prozorov}
\email[corresponding author:]{prozorov@ameslab.gov}
\affiliation{Ames National Laboratory, Ames, Iowa 50011, USA}
\affiliation{Department of Physics \& Astronomy, Iowa State University, Ames, Iowa 50011, USA}

\author{Makariy A. Tanatar}
\affiliation{Ames National Laboratory, Ames, Iowa 50011, USA}
\affiliation{Department of Physics \& Astronomy, Iowa State University, Ames, Iowa 50011, USA}

\author{Marcin Ko\'{n}czykowski}
\affiliation{Laboratoire des Solides Irradi\'{e}s, CEA/DRF/lRAMIS, \'{E}cole Polytechnique, CNRS, Institut Polytechnique de Paris, F-91128 Palaiseau, France}

\author{Romain Grasset}
\affiliation{Laboratoire des Solides Irradi\'{e}s, CEA/DRF/lRAMIS, \'{E}cole Polytechnique, CNRS, Institut Polytechnique de Paris, F-91128 Palaiseau, France}

\author{Alexei E. Koshelev}
\affiliation{Department of Physics and Astronomy, University of Notre Dame, Notre Dame, Indiana 46656, USA}

\author{Linlin Wang}
\affiliation{Ames National Laboratory, Ames, Iowa 50011, USA}
\affiliation{Department of Physics \& Astronomy, Iowa State University, Ames, Iowa 50011, USA}

\author{Sergey L. Bud'ko}
\affiliation{Ames National Laboratory, Ames, Iowa 50011, USA}
\affiliation{Department of Physics \& Astronomy, Iowa State University, Ames, Iowa 50011, USA}

\author{Paul C. Canfield}
\affiliation{Ames National Laboratory, Ames, Iowa 50011, USA}
\affiliation{Department of Physics \& Astronomy, Iowa State University, Ames, Iowa 50011, USA}

\date{1 June 2024}

\begin{abstract}
We report an unusual anisotropic paramagnetic peak effect observed in reversible magnetization of a single crystalline nodal superconductor $\text{Rh}_{17}\text{S}_{15}$. Both temperature- and field-dependent magnetization measurements reveal a distinct novel vortex state above approximately 1 T. This peak effect is most pronounced when the magnetic field, $H$, is applied parallel to the $\left[111\right]$ direction, whereas it diminishes for $H\parallel\left[110\right]$. Intriguingly, for $H\parallel\left[100\right]$, instead of a peak, we observe a step-like decrease in $M(T)$, with the step amplitude increasing in larger applied magnetic fields. This behavior is opposite to the expectations of conventional Meissner expulsion. The magnitude of the peak effect, expressed in terms of dimensionless volume susceptibility, is on the order of $\Delta\chi=10^{-5}$ (with full diamagnetic screening corresponding to $\chi=-1$). The observed anisotropic paramagnetic vortex response is unusual considering the cubic symmetry of $\text{Rh}_{17}\text{S}_{15}$. We propose that in this distinct vortex phase, a small but finite attractive interaction between vortices below $H_{c2}$ may be responsible for this unusual phenomenon. Furthermore, the vortices seem to prefer aligning along the $\left[111\right]$ direction, rotating toward it when the magnetic field is applied in other directions. Our findings add another item to the list of unusual properties of $\text{Rh}_{17}\text{S}_{15}$ that attracted recent attention as the first unconventional superconductor that has a mineral analog, miassite, found in nature. 
\end{abstract}
\maketitle

\section{Introduction}

Since the division of superconductors into two types by Abrikosov in 1957 \cite{Abrikosov1957}, vortex physics has become one of the most active research areas in superconductivity due to its rich fundamental physics aspects and direct relevance for technological applications
\cite{Pippard1969,Campbell1972,DewHuges1974,Larkin1979,Blatter1994,Brandt1995,Yeshurun1996,Hull2003,Tinkham2004}. Various vortex lattice phases and behaviors, usually mapped on the vortex $H-T$ phase diagram, are believed to be linked to the unconventional nature of superconductivity in different classes of superconductors, such as high$-T_{c}$ cuprates \cite{Blatter1994,Yeshurun1996,Hull2003,Kwok2016},
borocarbides \cite{Budko2006,Mueller2001} and iron-pnictides \cite{Canfield2010,Wen2011,Stewart2011,Chubukov2015,Kwok2016}.
One of the most interesting mixed-state features is the non-monotonic
dependence of the irreversible component of magnetization on a magnetic field or temperature. Depending on the context and author's preferences, this feature can be called the ``peak effect'', the ``second magnetization peak'' or the ``fishtail'' \cite{DeSorbo1964,Larkin1979,Daeumling1990,KrusinElbaum1992,Klein1994,Blatter1994,Brandt1995,Tang1996,Giller1997,Banerjee2000,Mikitik2001,Prozorov2008,Naren2009,Gao2022}.
Since any measurement has a certain experimental time window, the measured
magnetic moment or current density is affected by magnetic relaxation,
which is exponentially fast at current densities close to the critical
current \cite{Blatter1994,Brandt1995,Yeshurun1996}. Therefore, there
is always a question as to whether the peak effect is due to the actual
non-monotonic behavior of $j_{c}\left(H,T\right)$, which would imply
an unusual pinning mechanism, or whether it is the result of a non-monotonic
magnetic relaxation. This latter ``dynamic'' mechanism is predicted
in the weak collective pinning and creep model \cite{Blatter1994}
and the former ``static'' mechanism can be, for example, due to the softening of the vortex lattice at low fields and close to $H_{c2}$ \cite{Pippard1969,Larkin1979}, two different vortex phases \cite{Konczykowski2000} and a crossover from the collective to the plastic creep mechanism \cite{Abulafia1996,Giller1997}. Importantly, in either case, the reversible magnetization is always considered to be a monotonic function of a magnetic field and temperature. In fact, we are not aware of any reports of the non-monotonic behavior of reversible magnetization, except the so-called
paramagnetic Meissner effect (PME), which seems to be related to extrinsic
factors such as sample inhomogeneities, granularity, demagnetizing factors, form factor, or inhomogeneous cooling conditions \cite{Koshelev1995,Nusran2018,Koblischka2023}.

Superconducting $\text{Rh}_{17}\text{S}_{15}$ with $T_{c}=5.4\:\text{K}$
has recently attracted attention as the rare case of a cubic compound with line nodes in its superconducting gap inferred from the $T-$linear variation of the London penetration depth measured down to 50 mK, and a strong suppression of $T_c$ by non-magnetic disorder \cite{Kim2024}. Thermal conductivity measurements down to 100 mK on the lower $T_c=5.0\:\text{K}$ sample show nodal-like concave field dependence but no residual term \cite{Nie2024}. 

Despite its modest $T_{c}$, $\text{Rh}_{17}\text{S}_{15}$ exhibits quite unusual superconducting properties. It has a very large upper critical field, $H_{c2}\left(0\right)\approx20.5\:\text{T}$, determined from the fit to Helfand and Werthamer theory \cite{Helfand1966} of the experimental zero resistivity data measured along the $\left[111\right]$ direction where the last data point from the $R(H)$ scan at 0.5 K shows $H_{c2}$ just below 20 T \cite{Settai2010}. Data from polycrystalline samples yielded a similar value of $H_{c2}\left(0\right)\approx20\:\text{T}$ \cite{Naren2011,Naren2011b}. This $H_{c2}\left(0\right)$ exceeds the Pauli paramagnetic limiting field by a factor of two, $H_{p}(0)=\Delta(0)/(\sqrt{2}\mu_{B})\approx10\:\text{T}$, where the superconducting gap is estimated from the weak-coupling isotopic Bardeen-Cooper-Schrieffer (BCS) theory \cite{Bardeen1957}, $\Delta(0)=1.76k_{B}T_{c}\approx0.82\:\text{meV}$. There is a large difference between the coherence length of about $\xi\left(0\right)\approx4.0$~nm, derived from $H_{c2}(0)=\phi_{0}/(2\pi\xi^{2})$, and the BCS length scale $\xi_{0}=\hbar v_{F}/(\pi\Delta(0))$$\approx21.8\:\text{nm}$, which is more than five times greater. Here, $v_{F}=0.85\times10^{5}\:\text{m}/\text{s}$ is the band-averaged Fermi velocity evaluated by our DFT calculations. In the weak-coupling BCS theory, in the clean limit (which is the case here), these two lengths are of the same order, $\xi_{0}/\xi=1.63$ for isotropic $s-$wave \cite{Werthamer1966,Kogan2012} and, similarly, with a slightly different numerical prefactor, for arbitrary $k-$ dependent order parameter, including line nodal $d-$wave \cite{Benfatto2002}. Therefore, an extremely high $H_{c2}(0)$ alone represents a significant departure from the BCS theory for this relatively low$-T_{c}$ superconductor. All these properties imply that $\text{Rh}_{17}\text{S}_{15}$ is the first unconventional superconductor whose formula can be found in nature in the form of mineral miassite \cite{Kim2024}.

There is only limited information on the vortex properties of $\text{Rh}_{17}\text{S}_{15}$, because most studies have been performed on polycrystalline samples. For single crystals, only critical fields were reported, but without direction-resolved data. The broad peak effect in magnetization was observed as a function of the magnetic field and some non-monotonic signatures were observed in magnetic susceptibility as a function of temperature \cite{Naren2009}. The authors found a strong dependence of the persistent current density on the time window of the experiment and suggested a mix of static and dynamic scenarios where the vortex lattice becomes progressively more disordered with an increasing magnetic field. Perhaps, because of its cubic crystal structure, it was believed that this does not make much difference. However, cubic materials can exhibit electronic anisotropy \cite{Zarea2023}, including in the vortex state, for example of the upper critical field of niobium \cite{Butler1980}.

In this paper, we report the significant anisotropy of the vortex response in single crystals of $\text{Rh}_{17}\text{S}_{15}$ below the upper critical field, $H_{c2}$. A highly unusual paramagnetic \textit{reversible} peak effect was observed in magnetization as a function of temperature and magnetic field for $H\parallel\left[111\right]$ and it is reduced when $H\parallel\left[110\right]$. For $H\parallel\left[100\right]$ instead of a peak, a step in $M(T)$ develops. These obsevations indicate that vortices prefer to align along the $\left[111\right]$ ``easy direction''. The effect is suppressed by disorder. It appears that there is a novel vortex lattice phase that has never been reported before.

\section{Results}

The basic properties and characterization of our single crystals with a focus on the superconducting gap structure have previously been reported \cite{Kim2024}. In low fields, there is a sharp superconducting transition in magnetization and resistivity. In small magnetic fields, there is a significant difference between zero-field-cooled (ZFC) and field-cooled (FC) magnetization. For example, following the ZFC protocol by applying $H=\unit[10]{Oe}$ after cooling to a base temperature, $T=\unit[2]{K}$, results in volume magnetic susceptibility, $\chi\approx-1$, indicating a complete diamagnetic screening in all three orientations. Cooling in the same field from above $T_c$ gives only $\chi\approx -0.018,\,-0.023,\,-0.027$ in the $\left[100\right]$, $\left[110\right]$ and $\left[111\right]$ directions, respectively. This indicates that the pinning is weakest in the $\left[111\right]$ direction. With an increasing magnetic field, the transition curves smear and at $H=1.5\:\text{T}$ become reversible, showing practically no hysteresis between ZFC and FC magnetization. 

There is a small but notable paramagnetic signal in the normal state, also reported previously in polycrystalline samples \cite{Naren2011,Naren2011b}. Just above $T_c$, the volume magnetic susceptibility is on the order of $\chi=1\times 10^{-4}$. This is comparable to the magnetic susceptibility of tungsten at room temperature, which is regarded as ``non magnetic''. Interestingly, this paramagnetic signal is temperature dependent. It was suggested that sharp peaks in the density of states at the Fermi level could result in temperature-dependent Pauli spin susceptibility. This mechanism has also been suggested for $\text{V}_3\text{Si}$ where the paramagnetic susceptibility per vanadium ion is of the same order as for rhosium in $\text{Rh}_{17}\text{S}_{15}$ \cite{Labbe1967,Naren2011b}. Perhaps, this active magnetic channel is related to the peak effect reported here.

In discussing magnetic anisotropy, it is important to analyze the possible effect of the sample shape, which enters the magnetic response via demagnetizing factors $N$ \cite{Prozorov2018}. For our fields of interest, above 1 T, demagnetizing correction can be safely neglected. Specifically, the effective magnetic field strength, seen by the sample, is $H_{\text{eff}}=H_{\text{app}}-NM/V$, where $H_{\text{app}}$ is the applied magnetic field and  $V=1.7\times10^{-9}\:\text{m}^3$ is the volume of the sample. Even if we take an exaggerated value of the measured magnetic moment, $M=10^{-3}\:\text{erg/G}=10^{-6}\:\text{Am}^2$, the corresponding volume magnetization is $M/V=580\:\text{A/m}$. The applied magnetic field strength corresponding to a magnetic induction of 1 T is $H_{app}=B_{app}/\mu_0\approx8\times10^5\:\text{A/m}$. Since the demagnetizing factor is bound by $0\leq N <1$ (in our case $N\approx 0.35-0.55$ depending on the orientation), the $NM/V$ correction term is of the order of $200-400\:\text{A/m}$ ($2.5-5.0\:\text{G}$), which is more than three orders of magnitude smaller than the applied field strength ($10000\:\text{G}$).

\begin{figure}[tb]
\includegraphics[width=8.6cm]{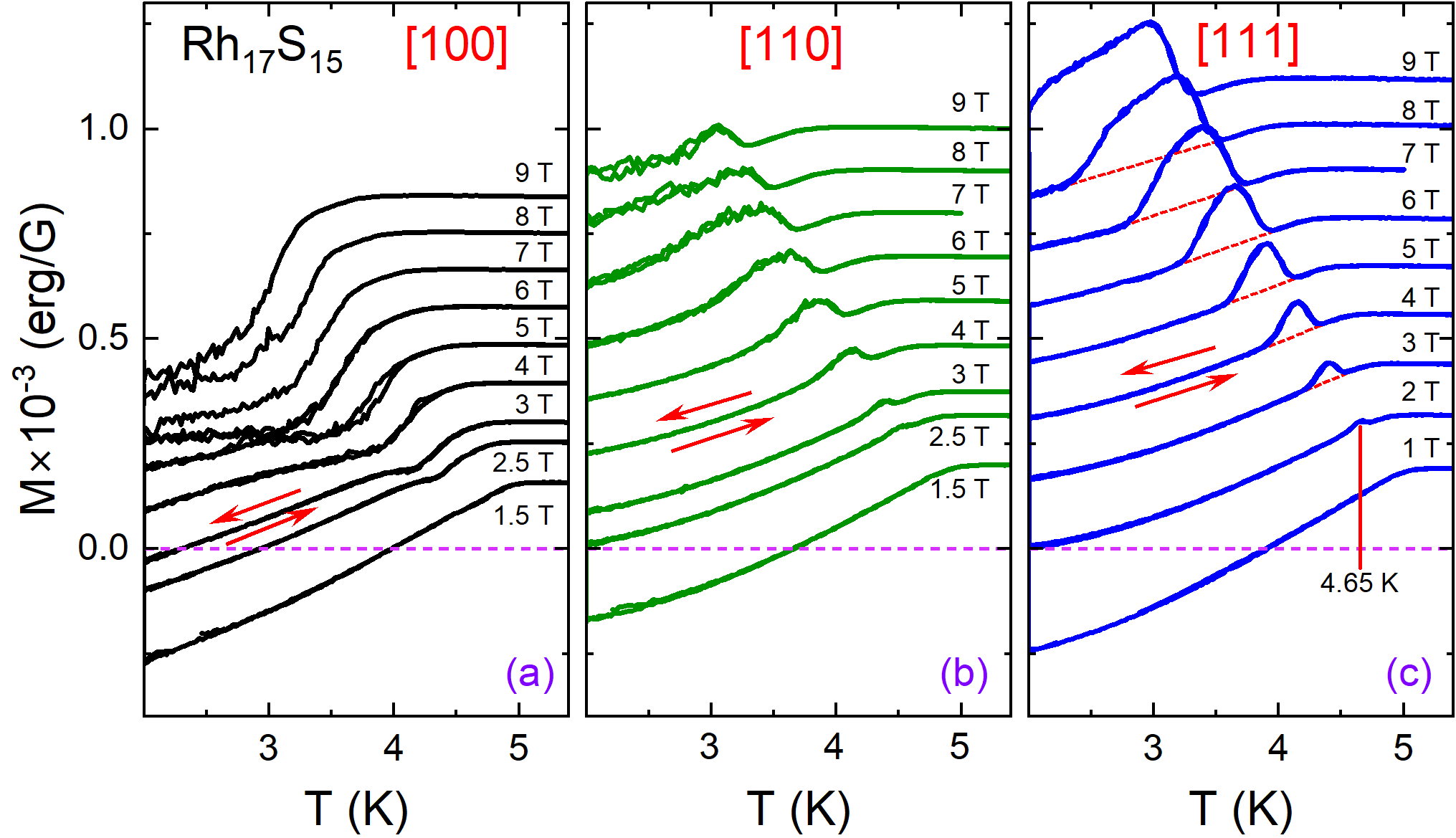}
\caption{
\label{fig1:M(T)-three-orientations} 
Zero-field-cooled (ZFC) and field-cooled (FC) magnetization of the same single crystal $\text{Rh}_{17}\text{S}_{15}$ in three different orientations, (a) $H\parallel\left[100\right]$, (b) $H\parallel\left[110\right]$, (c) $H\parallel\left[111\right]$. Each curve was measured in an indicated magnetic field and contains data on warming (ZFC) and cooling (FC), as shown by red arrows. The ZFC and FC curves coincide within the noise, indicating practically reversible magnetization. A pronounced reversible peak effect develops above $H=1.5\:\text{T}$ in $\left[111\right]$ orientation. Red dashed lines show that without the peak, there would be a continuous transition starting at $H_{c2}$. The peak is reduced in the $\left[110\right]$ direction. Instead of a peak, $\left[100\right]$ direction shows a step-like diamagnetic response with the step amplitude increasing with the increase of an applied magnetic field, unexpected for conventional Meissner expulsion.}
\end{figure}

Figure~\ref{fig1:M(T)-three-orientations} shows the ZFC and FC magnetization
of the same $\text{Rh}_{17}\text{S}_{15}$ single crystal in three different
orientations, (a) $H\parallel\left[100\right]$, (b) $H\parallel\left[110\right]$, and (c) $H\parallel\left[111\right]$, measured in indicated magnetic fields. Each curve contains the ZFC and FC data, which are indistinguishable within the experimental noise, as shown by red arrows. The magnetization is practically reversible. Note that we chose not to convert magnetization (shown as raw data in CGS erg/G units) to avoid division by a magnetic field, which would make the graph less clear.

The central result of this paper is an unexpected pronounced, reversible paramagnetic peak developing roughly above $H=1.5\:\text{T}$ in a $\left[111\right]$ orientation, but not present in the $\left[100\right]$ direction where there is a pronounced, steplike diamagnetic decrease in $M(T)$ on cooling, but no peak. In the intermediate orientation, $\left[110\right]$, there is a reduced peak effect. The noise developing below the transition in Fig.~\ref{fig1:M(T)-three-orientations} (a) and (b) is not due to instrumentation but is generated by the sample and most likely reflects the enhanced motion of vortices. The red dashed lines in Figure~\ref{fig1:M(T)-three-orientations}(c) show that without the peak, there would be a continuous transition starting at $H_{c2}$. This indicates that a new distinct vortex phase is formed inside the superconducting phase, disconnected from the $H_{c2}(T)$ line.

The peak in a $\left[111\right]$ direction is truly paramagnetic, as it exceeds the value in the normal state above $T_c(H_{c2})$, where the dimensionless paramagnetic volume susceptibility is of the order of $\chi=1\times 10^{-4}$, and the peak height from the baseline is approximately ten times smaller; see the inset in Fig.\ref{fig6:phase-dia}. We also stress that, while the $\left[100\right]$ direction does not show a peak, its temperature dependence is also unusual. There is a step in $M(T)$ below $H_{c2}$ and, importantly, its amplitude \textit{increases} with the increase of the applied magnetic field, contrary to the expectations of the standard Meissner expulsion. As can be seen in Fig.~\ref{fig1:M(T)-three-orientations}(a), there is practically no step in $M(T)$ at 1.5 T, but it grows with the applied field and its amplitude is comparable to the amplitude of the peak in the $\left[100\right]$ direction. 

\begin{figure}[tb]
\includegraphics[width=8.6cm]{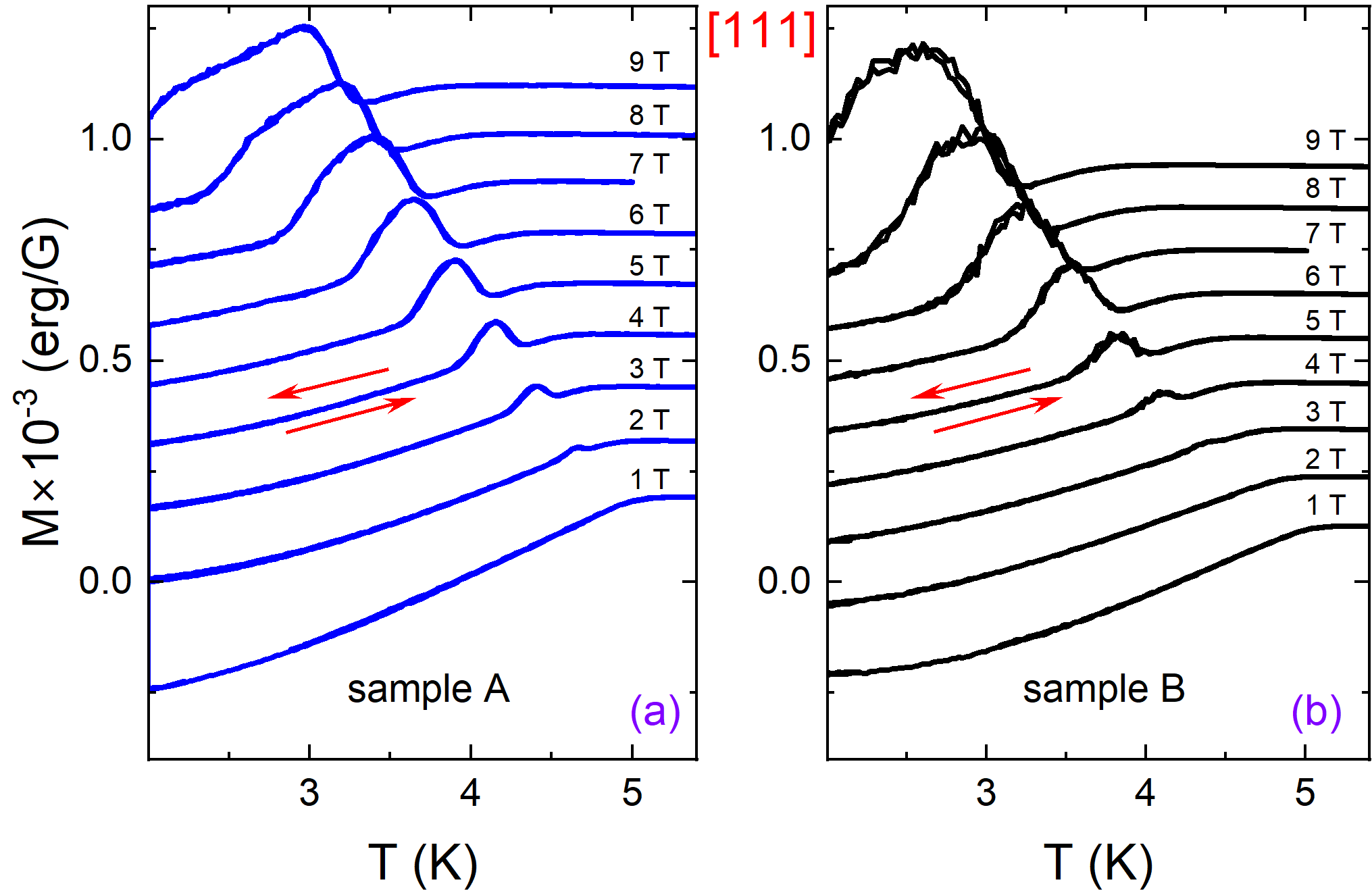}
\caption{
\label{fig2:M(T)-diff-samples} 
Zero-field-cooled (ZFC) and field-cooled (FC) magnetization of two different crystals of $\text{Rh}_{17}\text{S}_{15}$ measured with $H\parallel\left[111\right]$ in indicated magnetic fields. The curves are reversible as indicated by red arrows.}
\end{figure}

With such unusual observations, the first step is to verify the effect on other samples from a different batch. This is shown in Fig.~\ref{fig2:M(T)-diff-samples} where the temperature-dependent magnetic moment measured along the $\left[111\right]$ orientation is plotted for two different samples, (a) sample A and (b) sample B from a different batch. We measured several other samples and confirmed that the effect is reproducible.

\begin{figure}[tb]
\includegraphics[width=8.6cm]{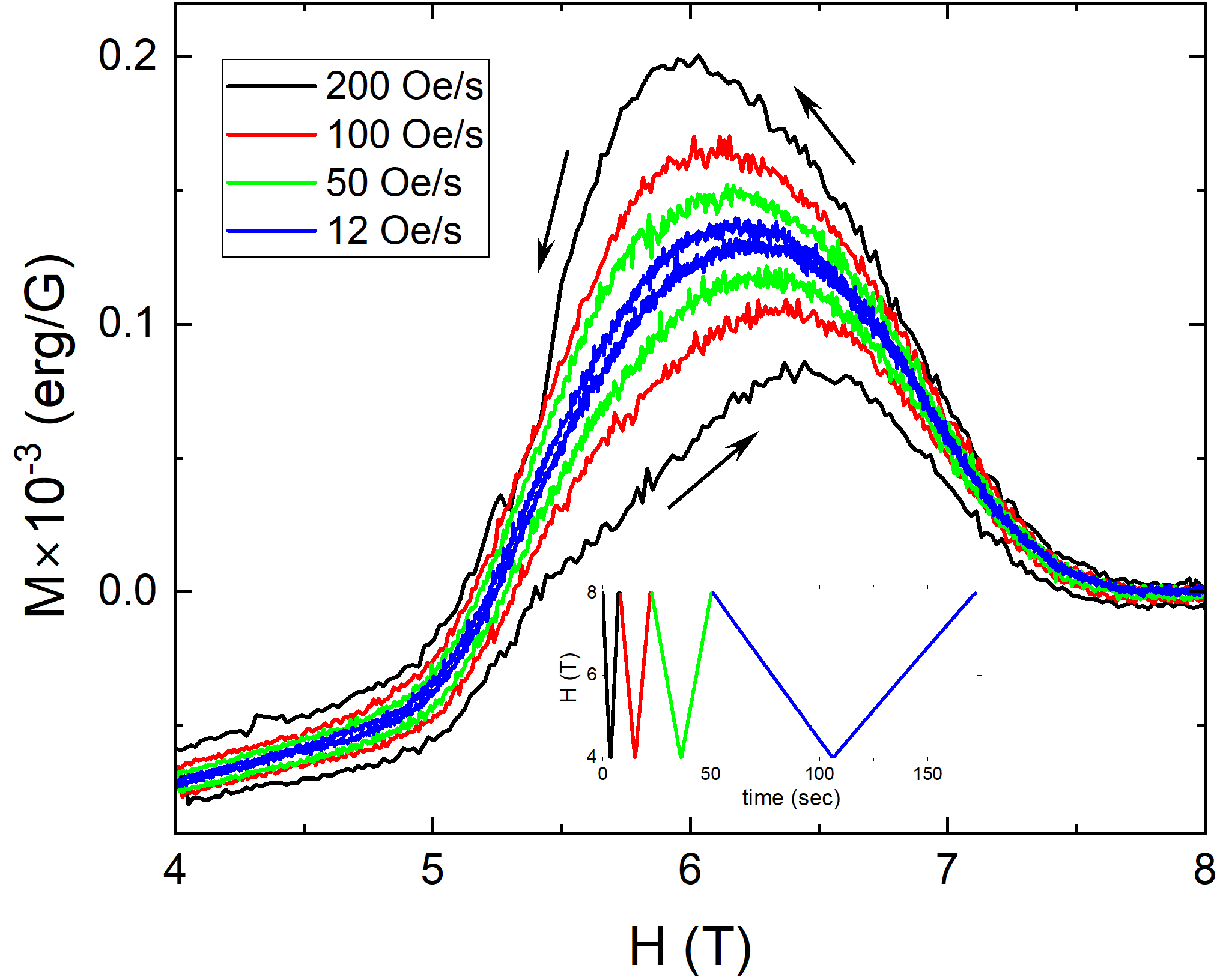}
\caption{
\label{fig3:M(H)-peak} 
Magnetic moment in sample A as a function of the applied magnetic field measured along the $\left[111\right]$ direction at $T=3.6\:\text{K}$. The up and down sweeps are performed at the same rate in the directions indicated by the arrows. Then the sweeps were repeated at different ramp rates indicated in the inset, which shows the time profile of the magnetic field sweeps. There is a significant magnetic relaxation with the blue curve obtained at $12\:\text{Oe/s}$ showing practically reversible $M(H)$ loop.}
\end{figure}

The next step is to examine the magnetic field dependence of the magnetic moment measured at a fixed temperature. Here we observed another surprising result, the peak effect, but
unlike the peak effects and fishtails reported so far, this peak effect
has a significant, dominant, reversible component, which is consistent with the peak shown in Fig.~\ref{fig1:M(T)-three-orientations}. Figure \ref{fig3:M(H)-peak} shows the magnetic moment measured along the $\left[111\right]$ direction at $T=3.6\:\text{K}$ in sample A. The external magnetic field was swept as shown in the inset in Fig.~\ref{fig3:M(H)-peak}, repeating down and up sweeps at the same rate, indicated by arrows, and then switching the ramp rate and repeating the measurement. Four different ramp rates, from 200 Oe/s, 100 Oe/s, 50 Oe/s and 12 Oe/s, were used, a seventeen-fold reduction. There is a significant magnetic relaxation with the blue curve obtained at the slowest 12 Oe/s showing a practically reversible nonmonotonic $M(H)$ loop. 

\begin{figure}[tb]
\includegraphics[width=8.6cm]{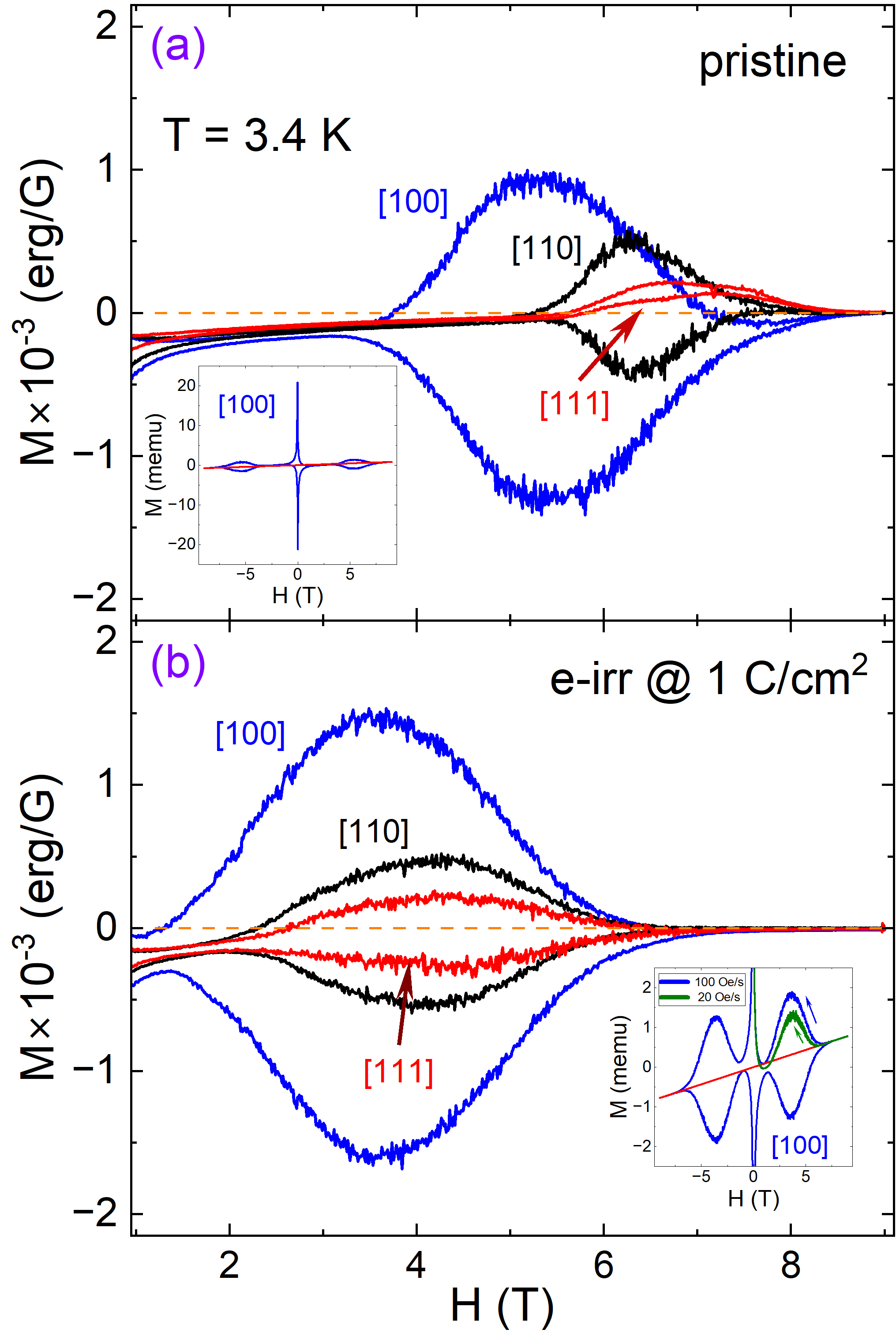}
\caption{
\label{fig4:M(H)-three-orientations} 
Magnetic hysteresis $M(H)$ loops of the same single crystal of $\text{Rh}_{17}\text{S}_{15}$ measured in sample A along three principal directions at $T=3.4\:\text{K}$ with small paramagnetic background subtracted. The magnetic field ramp rate was 100 Oe/s. Panel (a) shows the results in the pristine crystal and panel (b) shows the same crystal but after $2.5\:\text{MeV}$ electron irradiation with the dose of $6.2\times10^{18}\:\text{electrons}/\text{cm}^{2}$. The $x-$ and $y-$ axes scales are shared in both panels to allow for direct comparison. Insets show $M(H)$ loops before background subtraction: (a) a full loop including a pronounced zero-field peak; (b) a truncated $M(H)$ loop shown to emphasize the background line.}
\end{figure}

We now compare the magnetic hysteresis $M\left(H\right)$ loops measured
in the same sample A in three principal orientations. Figure \ref{fig4:M(H)-three-orientations}(a) shows the significant anisotropy of the response. There is a conventionally looking ``fishtail'' in the $\left[100\right]$ orientation (blue curve), but an asymmetric, almost reversible, peak effect in the $\left[111\right]$ orientation (red curve). Estimated from the width of the magnetic hysteresis, the maximum persistent current for the $\left[100\right]$ orientation is about ten times greater than for the $\left[111\right]$ orientation, and is intermediate for the $\left[110\right]$ orientation. The location of the peak shifts to the lower fields from $\left[111\right]$ to $\left[110\right]$  to  $\left[100\right]$. Such significant anisotropy in an electronically isotropic cubic system is remarkable but apparently not impossible in the vortex state. In order to better compare the loops, a small paramagnetic background, linear in $H$, was subtracted. The original loops are shown in the insets with backgrounds shown by red lines. In the pristine state, a full $M(H)$ loop is shown, including an often-observed sharp peak near zero field. In this scale, the background is practically zero. For the irradiated sample, discussed later, a truncated loop is shown to better visualize the background contribution. Importantly, the paramagnetic background is practically identical for all three orientations as well as before and after the electron irradiation.

Another important metric of a superconductor is the response to a controlled disorder. Previously, we used electron irradiation to probe the superconducting transition temperature of Rh$_{17}$S$_{15}$ and found results consistent with the line-nodal superconducting gap \cite{Kim2024}. Here we examine the effect of non-magnetic pointlike disorder on the observed paramagnetic reversible peak effect. Figure \ref{fig4:M(H)-three-orientations}(b) shows magnetization loops in the peak region for three orientations after $2.5\:\text{MeV}$ electron irradiation with the dose of $6.2\times10^{18}\:\text{electrons}/\text{cm}^{2}$. For direct comparison with the pristine case, the $x-$ and $y-$ axes scales are shared in both panels (a) and (b) of Fig.~\ref{fig4:M(H)-three-orientations}, pristine and irradiated, respectively. The initially almost reversible peak in the $\left[111\right]$ orientation evolves with disorder into a ``normal'' fishtail dominated by an irreversible component. Considering the very low pinning in our $\text{Rh}_{17}\text{S}_{15}$ crystals, it is possible that the peak effect observed in other materials with larger pinning may hide the reversible phase in their hypothetical clean state.

\begin{figure}[tb]
\includegraphics[width=8.6cm]{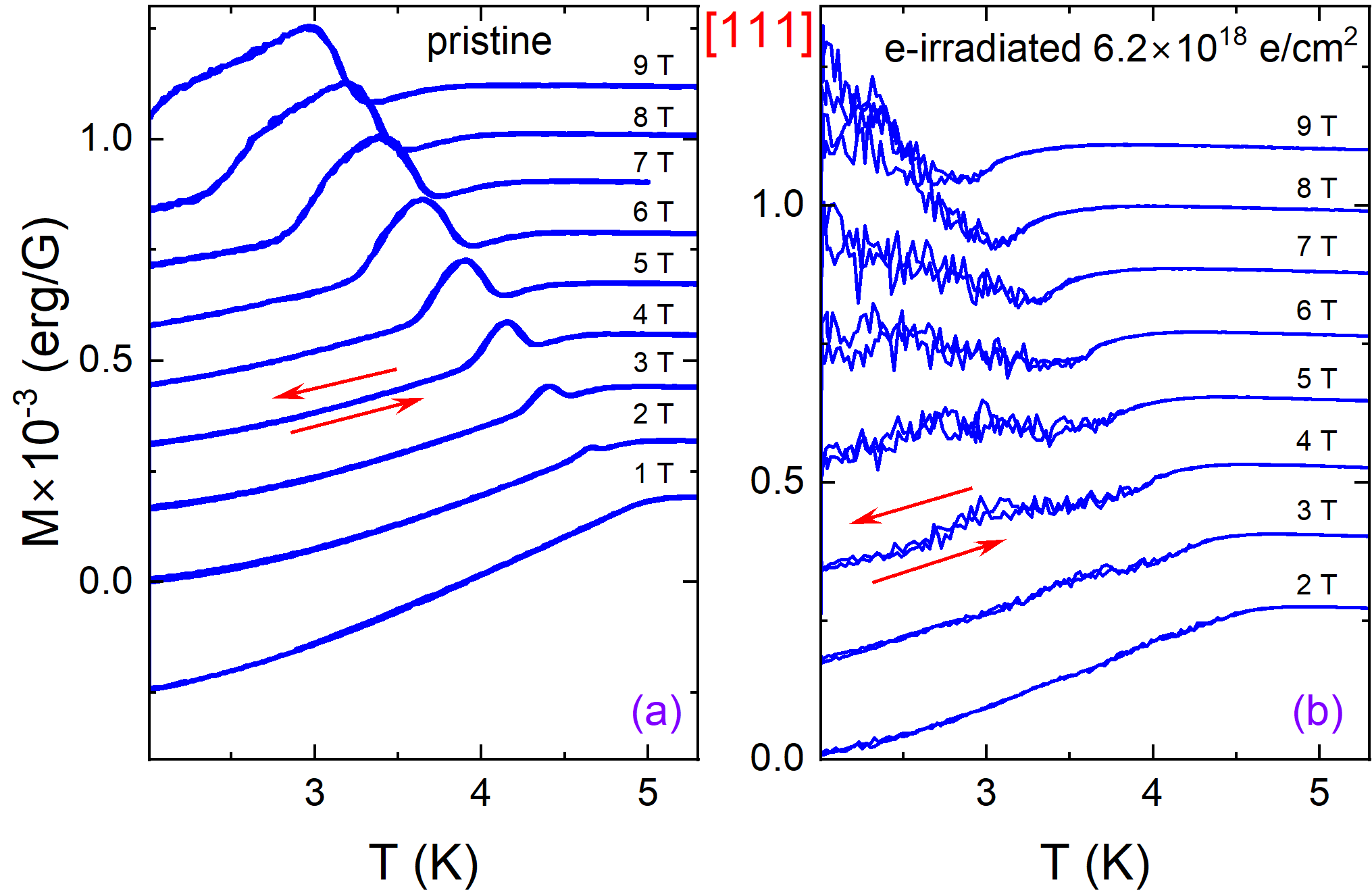}
\caption{
\label{fig5:M(T)-irr}
ZFC and FC magnetization of the same $\text{Rh}_{17}\text{S}_{15}$ single crystal A measured along $\left[111\right]$ direction, (a) before and (b) after $2.5\:\text{MeV}$ electron irradiation with the dose of $6.2\times10^{18}\:\text{electrons}/\text{cm}^{2}$.}
\end{figure}

\begin{figure}[tb]
\includegraphics[width=8.6cm]{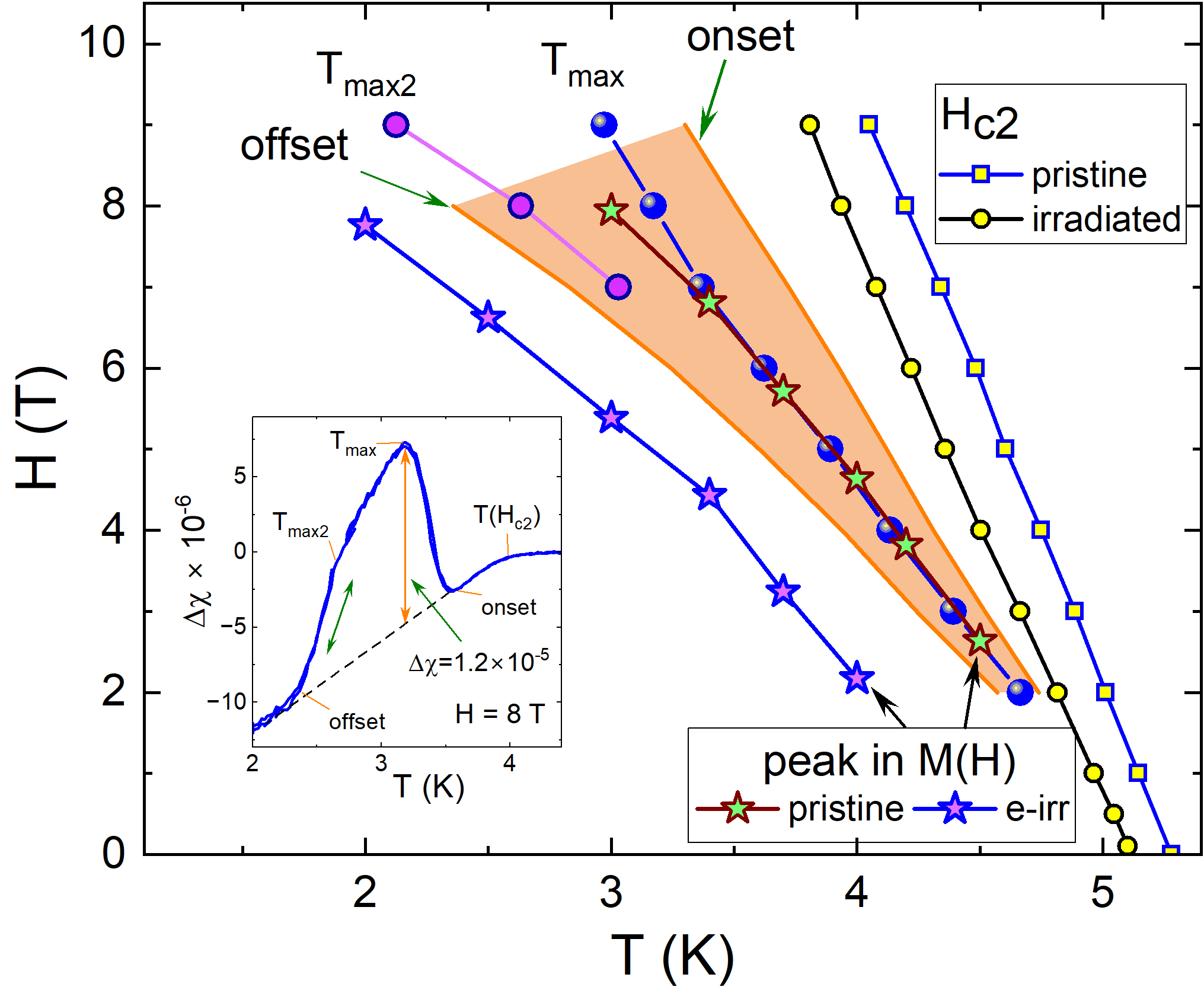}
\caption{
\label{fig6:phase-dia} 
The $H-T$ state phase diagram including the novel vortex phase. The definitions of characteristic temperatures of the reversible $M\left(T\right)$ peak are shown in the inset. The blue circles mark the temperature of the peak maximum $T_{max}$, and the violet circles show the temperature of the knee (can also be seen as the second maximum) of the peak, $T_{max2}$. Onset and offset temperatures are the boundaries of the orange area inside which the peak is located. Green stars and purple stars mark the location of the peak effect maximum in the $M\left(H\right)$ measurements in pristine and electron-irradiated samples, respectively. The yellow squares and circles show the upper critical field in a $\left[111\right]$ direction before and after electron irradiation, respectively. In the inset, the paramagnetic background above $T_c$ was subtracted to show the volume susceptibility of the peak in absolute units. 
}
\end{figure}

The destructive effect of disorder on the observed reversible peak effect is further illustrated in Fig.~\ref{fig5:M(T)-irr}, where $M(T)$ ZFC-FC temperature scans are compared for the same sample A in a $\left[111\right]$ orientation in (a) pristine and (b) electron irradiated states. The peak is smeared and, again, magnetic noise appears similar to that in Fig.~\ref{fig1:M(T)-three-orientations}(a),(b), probably indicative of significant vortex displacements and jumps between metastable configurations.

Finally, we construct an $H-T$ phase diagram that includes the novel reversible vortex phase. Figure \ref{fig6:phase-dia} shows different $H(T)$ lines of characteristic temperatures defined as shown in the inset. In the inset, the paramagnetic background above $T_c$ was subtracted to show the paramagnetic volume susceptibility of the peak in absolute units. The blue circles mark the temperature of the peak maximum, $T_{max}$, and violet circles show the peak shoulder, $T_{max2}$. The orange area is bound by the onset and offset temperatures. It represents the domain inside which the novel vortex phase exists in our range of fields and temperatures. For comparison, additional lines are obtained tracing the peak position in the $M\left(H\right)$ loops measured at different temperatures. Green- and violet-filled stars show the results for the pristine and irradiated samples, respectively. Remarkably, the peak positions determined from the two measurements coincide in pristine samples. There is no peak in the $M(T)$ measurements after irradiation and the peak in the $M(H)$ loops is shifted to lower temperatures.
To complete the phase diagram, Fig.~\ref{fig6:phase-dia} shows the upper critical field in a $\left[111\right]$ orientation before (yellow squares) and after (yellow circles) electron irradiation. The line $H_{c2}(T)$ is shifted to lower temperatures after irradiation, consistent with our earlier conclusion that $\text{Rh}_{17}\text{S}_{15}$ is a line nodal superconductor.  

\section{Discussion}

Although we do not have a microscopic explanation for the unusual anisotropic reversible vortex behavior, we can state that we observed a distinct intermediate vortex state likely related to another order which may be either magnetic or from a competing superconducting channel. Due to the reversible nature of the ZFC-FC magnetization and the tendency to such a behavior in $M(H)$ loops after relaxation of the irreversible component, we can exclude non-equilibrium flux trapping and vortex density gradients associated with pinning. This implies that a uniform vortex lattice experiences some compression, which means that the vortices attract each other. Despite a very low pinning the field-cooled magnetic susceptibility is surprisingly small, which may also indicate that the Meissner expulsion is hindered by an attractive force. 

The observed paramagnetic effect is small. In terms of dimensionless volume magnetic susceptibility, the peak height increases with increasing magnetic field and appears to saturate around $\Delta\chi\approx+10^{-5}$. This is on top of the paramagnetic background above $T_c$, which is about $\chi\approx+10^{-4}$. We can estimate how much the distance between the vortices should change to cause such a peak. For a triangular lattice, the intervortex distance is, $a=\sqrt{2\phi_{0}/\sqrt{3}B}$. The magnetic susceptibility for uniform $B$ is, $\chi=\left(B/H_{app}-1\right)$
(as discussed above, the demagnetizing correction is not important in large fields of the effect). Therefore, to the first order, the
change in the intervortex spacing, $\Delta a=a-a_{0}\approx-\Delta\chi a_{0}/2$. Positive $\Delta\chi$ means compression. Numerically, for $\mu_0H_{app}=9\:\text{T}$, the spacing is $a_{0}=16.3\:\text{nm}$ and we find that the vortices must move by a small distance, $\Delta a\approx9.7\times10^{-5}\:\text{nm}$. Therefore, if some attractive force causes the observed paramagnetic peak, it is extremely small yet finite. An interesting and important aspect is that the situation is highly anisotropic with a step instead of a peak observed in the $\left[100\right]$ direction. To explain this, we suggest that the vortices tend to alight along the $\left[111\right]$ crystallographic direction, so that when a magnetic field is aligned in other directions and in the absence of pinning, vortices rotate toward $\left[111\right]$ causing a reduction of the $\left[100\right]$ component of the magnetization vector. Since our magnetic measurements are always performed along the direction of an applied magnetic field, this leads to a step-like feature in $\left[100\right]$ orientation.

Throughout the paper, we emphasized a small but finite paramagnetic background. Considering the paramagnetic nature of the described effect, it is possible that normal-state magnetism plays some role, although the mechanism is unclear.

In the discussion of the unusual effects observed, it is important to estimate some relevant parameters for $\text{Rh}_{17}\text{S}_{15}$. First, we evaluate the fundamental length scales. Using the thermodynamic Rutgers relation, we previously obtained the value of London penetration depth of $\lambda\left(0\right)=550\:\text{nm}$ \cite{Kim2024}. The only other work on single crystals used magnetization measurements to determine the lower critical field, which yielded a similar value of $\lambda\left(0\right)=490\:\text{nm}$ \cite{Settai2010}, although such estimates have a very large error bar. 
The coherence length is $\xi(0)=4\:\text{nm}$ for $H_{c2}(0)=20.5\:\text{T}$. Therefore, the Ginzburg-Landau parameter $\kappa\approx137.3$ places $\text{Rh}_{17}\text{S}_{15}$ in an extreme type-II limit. Overall, single crystals show a very small amount of hysteresis with reversible magnetization dominating over the irreversible part, even in the regions of the ``fishtail'' in $M\left(H\right)$ loops. This is the case for all orientations of the applied magnetic field. However, in the peak effect region, the hysteresis in the $\left[111\right]$ orientation is about ten time smaller than in the $\left[100\right]$ orientation. This clean-limit behavior is also consistent with the nearly perfect $T-$linear behavior of $\lambda\left(T\right)$ at low temperatures \cite{Kim2024}. Furthermore, from the resistivity at the transition temperature, $\rho(T_{c})\approx12\;\mu\Omega\cdot\text{cm}$, the mean free path can be estimated using the electronic band-structure parameters, $\ell=3/(v_{F}N\left(0\right)e^{2}\rho)\approx 91\:\text{nm}$,
where the band-averaged product $v_{F}N\left(0\right)=1.1\times10^{52}\:\text{s}^{-1}\text{m}^{-2}\text{J}^{-1}$. This yields the dimensionless scattering rate, $\Gamma=0.88\xi_{0}/\ell=0.2<1$, placing the material in the clean limit. 

Vortex-relevant parameters include: the lower critical field, $H_{c1}\left(0\right)=\phi_{0}\left(\ln \kappa+0.497\right)/(4\pi\lambda^{2})\approx29\:\text{G}$
for $\lambda(0)=550\:\text{nm}$ and $\xi(0)=4\:\text{nm}$; the characteristic
energy scale, $\varepsilon_{0}=\phi_{0}^{2}/\left(4\pi\mu_{0}\lambda^{2}(0)\right)=5.6\:\text{meV}/\text{nm}$; and vortex line energy, $\varepsilon_{l}=\varepsilon_{0}(\ln\kappa+0.497)=30.2\:\text{meV}/\text{nm}$.
The thermodynamic critical field, $H_{c}(0)=\phi_{0}/\left(2\sqrt{2}\pi\xi(0)\lambda(0)\right)=0.11\:\text{T}$
and the Ginzburg number that characterizes the width of the critical
fluctuations, $\ensuremath{Gi=\left(\gamma k_{B}T_{c}/\varepsilon_{0}\xi(0)\right)^{2}/8=5.3\times10^{-5}}$;
here we assumed anisotropy parameter $\gamma=1$. A practical formula
for this parameter is: $Gi=3.25\times10^{-16}[\text{nm}^{-2}\text{K}^{-2}]\left(\kappa\gamma\lambda(0)[\text{nm}]T_{c}[\text{K}]\right)^{2}$. The largest $Gi$ value is observed in high-$T_{c}$ cuprates, of the order of $10^{-2}$ or more. For low-$T_{c}$ conventional superconductors it is in the range of $<10^{-7}$ and much less. For example, in niobium, assuming the clean case values of $\lambda(0)=33\:\text{nm}$, $\xi(0)=93\:\text{nm}$ and $T_{c}=9.4\:\text{K}$ \cite{Prozorov2022}, $Gi=4\times10^{-12}$. In iron-based superconductors, with the typical values of $\lambda(0)=200\:\text{nm}$, $\xi(0)=3\:\text{nm}$ and $T_{c}=33\:\text{K}$ \cite{Prozorov2011}, we obtain $Gi=8\times10^{-5}$, see also \cite{Koshelev2019}. Therefore,  the Ginzburg number of $\text{Rh}_{17}\text{S}_{15}$ is comparable with iron pnictides and it is unusually large for low-$T_{c}$ superconductors.

\section{Conclusions}

We report an unusual highly anisotropic reversible vortex phase in single crystals $\text{Rh}_{17}\text{S}_{15}$. For $H\parallel\left[111\right]$, there is a pronounced peak in $M(T)$. This peak is reduced for $H\parallel[110]$. For $H\parallel\left[100\right]$, instead of a peak, a steep step develops. Both the peak and the step amplitudes increase with the increase of an applied magnetic field. In all orientations, the warming and cooling scans are reversible. For $H\parallel\left[111\right]$, there is also a reversible peak effect (inside the irreversible ``fishtail'') in $M(H)$ loops with the same location as in $M(T)$ scans when plotted on a $T-H$ phase diagram. Non-magnetic point-like disorder induced by electron irradiation suppresses the unusual features. We suggest that the observed peak effect may be caused by a weak attractive interaction between vortices. Furthermore, $H\parallel\left[111\right]$ appears to be the ``easy axis'' for vortices. When a magnetic field is applied in a different direction, the vortices rotate toward $H\parallel\left[111\right]$, leading to a reduction in the magnetization projection on the measurement axis, which explains the step observed in the $H\parallel\left[100\right]$ orientation.

\section{Methods}

\noindent \textbf{Single crystal growth:} 
Single crystalline samples of $\text{Rh}_{17}\text{S}_{15}$ were synthesized out of the Rh-S eutectic by using a high-temperature solution growth technique. The details of growth and characterization are provided elsewhere \cite{Canfield2023,Kim2024}. Briefly, elemental rhodium powder and sulfur were combined in a fritted Canfield Crucible set\cite{Canfield2016}, sealed in a silica ampoule, slowly heated (over 12 hours) to 1150\,\celsius~ and then slowly cooled from 1150\,\celsius~ to 920\,\celsius~ over 50 hours and decanted\cite{Canfield2019}. Millimeter-sized single crystals of $\text{Rh}_{17}\text{S}_{15}$ with well-defined facets were obtained.

\noindent \textbf{Samples:}
Cuboid-shaped samples with visible $\left(100\right)$, $\left(110\right)$ and $\left(111\right)$ facets were selected for measurements. Sample A, used for most figures and quantitative analysis, had dimensions: $1.35 \times 1.25\times 1.10\:\text{mm}^3$ along three orthogonal $\left<100\right>$ directions. It had volume $V=1.72\:\text{mm}^3$ and weighed $12.68\:\text{mg}$. Sample B was $1.35 \times 1.0\times 0.95\:\text{mm}^3$, $V=1.28\:\text{mm}^3$ and weight $9.43\:\text{mg}$. We measured several other samples for statistics and consistently observed reported here results.

\noindent \textbf{Magnetization measurements:} 
Magnetic moment was measured using a \textit{Quantum Design} vibrating sample magnetometer (VSM) in a $9\:\text{T}$ \textit{Physical Property Measurement System} (PPMS). Note that the magnetic moment is measured in CGS units, $1~ \text{emu} = 1~\text{erg/G}=10^{-3}\text{Am}^2$. The samples were glued in proper orientation on Pyrex cylinders fitted inside the brass half-cylinder, a standar sample holder suitable for our samples. The orientation was done by eye following well-defined crystallographic planes. The same sample was repeatedly oriented and measured, then irradiated, and then the same set of measurements was repeated. Other samples from different batches were measured.

\noindent \textbf{Electron irradiation:} 
The low-temperature 2.5 MeV electron irradiation was performed at the \textit{``SIRIUS''} facility of the \textit{Laboratoire des Solides Irradi\'{e}s} (LSI) at \textit{\'{E}cole Polytechnique} in Palaiseau, France. The general description of such experiments and specific details are provided elsewhere \cite{Damask1963,Thompson1969,Cho2018}. Relativistic electrons are particularly suitable for the creation of point-like defects in solids, because the energy transfer upon their collisions with ions matches the knockout energy barriers, typically in the $10-100\:\text{eV}$ range in metallic compounds. A knocked out ion becomes an interstitial and, together with the created vacancy, forms the so-called Frenkel pair. The irradiation is carried out with the sample immersed in liquid hydrogen at about $22\:\text{K}$ to avoid immediate recombination of Frenkel pairs and clustering of the produced defects. Upon warming to room temperature, a metastable population of defects remains after some defects recombined and migrated to surfaces and extended defects \cite{Damask1963,Thompson1969}. The degree of annealing depends on the material and in $\text{Rh}_{17}\text{S}_{15}$ is about 30\%. The amount of the produced disorder is gauged by the increase in resistivity after irradiation. Public domain software developed by NIST (https://physics.nist.gov) was used for electron propagation and energy loss calculations and custom SECTE software developed at the LSI at \textit{\'{E}cole Polytechnique} was used for the scattering cross-section calculations.
In this work, a single dose of $6.24\times10^{18}$ electrons/cm$^{2}$ was used. 
The total cross-section to knock out either Rh or S at 2.5 MeV is 76.4 barn, assuming a typical displacement threshold of 30 eV. This gives about one defect per approximately 30 conventional formula units ($Z=2$) with a mean distance between the defects of approximately 3 nm. The resistivity at $T_c$ increased from $12\:\mu\Omega\cdot\text{cm}$ to approximately $30\:\mu\Omega\cdot\text{cm}$, corresponding to a threefold increase in the scattering rate $\Gamma$, proportional to the concentration of defects. With the above estimate, $\Gamma=0.2$ in the clean case, and it increases to $\Gamma=0.6$ upon irradiation that moves closer to the dirty limit of $\Gamma>1$. The increase in pinning due to additional disorder is evident in the magnetization $M(H)$ loops presented.

\begin{acknowledgments}
This work was supported by the U.S. Department of Energy (DOE), Office of Science, Basic Energy Sciences, Materials Science and Engineering Division. Ames Laboratory is operated for the U.S. DOE by Iowa State University under contract DE-AC02-07CH11358. Electron irradiation was performed on the SIRIUS platform supported by the French National network of accelerators for irradiation and analysis of molecules and materials EMIR\&A (FR CNRS 3618). 
\end{acknowledgments}

%

\end{document}